\documentclass[twocolumn]{aastex63}

\shorttitle{DECALS Shear Measurement}
\shortauthors{Zhang et al.}

\usepackage{amsmath}
\begin{document}

\title{Shear Measurement with Poorly Resolved Images}

\correspondingauthor{Jun Zhang}
\email{betajzhang@sjtu.edu.cn}

\author{Jun Zhang}
\affiliation{Department of Astronomy, Shanghai Jiao Tong University, Shanghai 200240, China}
\affiliation{Shanghai Key Laboratory for Particle Physics and Cosmology, Shanghai 200240, China}

\author{Cong Liu}
\affiliation{Department of Astronomy, Shanghai Jiao Tong University, Shanghai 200240, China}

\author{Pedro Alonso Vaquero}
\affiliation{Department of Astronomy, Shanghai Jiao Tong University, Shanghai 200240, China}

\author{Hekun Li}
\affiliation{Department of Astronomy, Shanghai Jiao Tong University, Shanghai 200240, China}

\author{Haoran Wang}
\affiliation{Department of Astronomy, Shanghai Jiao Tong University, Shanghai 200240, China}

\author{Zhi Shen}
\affiliation{Department of Astronomy, Shanghai Jiao Tong University, Shanghai 200240, China}

\author{Fuyu Dong}
\affiliation{School of Physics, Korea Institute for Advanced Study (KIAS), 85 Hoegiro, Dongdaemun-gu, Seoul, 02455, Republic of Korea}

\begin{abstract}
Weak lensing studies typically require excellent seeing conditions for the purpose of maximizing the number density of well-resolved galaxy images. It is interesting to ask to what extent the seeing size limits the usefulness of the astronomical images in weak lensing. In this work, we study this issue with the data of the DECam Legacy Survey (DECaLS), which is a part of the target selection program for the Dark Energy Spectroscopic Instrument (DESI). Using the Fourier\_Quad shear measurement pipeline, we demonstrate that images with relatively poor seeing conditions ($\sim 1.5"$) can still yield accurate shear estimators. We do not find any correlation between systematic shear error and the image resolution.

\end{abstract}

\keywords{gravitational lensing: weak – large-scale structure of universe – methods: data analysis}

\section{Introduction} \label{sec:intro}

One of the key objectives of many ongoing and planned large scale galaxy surveys is to perform accurate measurement of the weak lensing effect, in the hope of resolving the standing tension about the constraints of the cosmological parameter $S_8$ \citep{Hildebrandt2017,Hikage2019,Joachimi2021,Abbott2022,Amon2022}, as well as probing the nature of dark matter \citep{Hoekstra2008,Uitert2017}, dark energy \citep{Dong2019}, and the theory of gravity \citep{Zhang2007,Simpson2013,Liu2016,Brouwer2017,Amon2018}.  

In such studies with ground-based telescopes, one typically requires good seeing conditions, so that a large fraction of galaxy shapes can be well resolved, and the smearing effect due to the point spread function (PSF) can be properly corrected \citep{Miller2013,Jee2013,Kuijken2015,Mandelbaum2018,Fu2018,Jarvis2021}. Implicit or explicit cuts on the galaxy size (relative to the PSF size) are often imposed in currently popular shear measurement methods to guarantee accuracy at certain levels \citep{Mandelbaum2005,Massey2013,Conti2017,Liu2018}. These requirements strongly limit the number of useful source images, and therefore the scientific potentials of galaxy surveys.

For galaxies that are much smaller than the PSF, it is perhaps not surprising that any attempt of measuring their shapes should bear significant amount of uncertainties. This is because the shape information in this case is all confined on small scales, which suffers from strong suppression by the PSF\footnote{It may help to think in Fourier space.}, and confusion by the noise. It is for this reason that most shear measurement methods require cuts on the galaxy size in practice. Interestingly, in our earlier work \citep{Zhang2016}, we find that such a cut is not neccessary in the Fourier\_Quad method (FQ hereafter). The reason is that in FQ, the shape uncertainties at the "small-size" end can be properly presented in the mathematical form of "zero-divide-zero", which turns out not to influence the overall accuracy of shear recovery. It has been shown that the lensing statistics are not affected even if point sources are included in the sample. This feature of FQ allows us to include many poorly resolved sources in the shear statistics. The purpose of this paper is to study the actual performance of FQ under very general observing conditions.

The data we use is from the DESI Legacy Imaging Surveys \citep{Dey2019}, which was designed mainly for the target selection of the following-up spectroscopic measurement, not for weak lensing. In these images, as we show later in the paper, the distribution of the Full-Width-at-Half-Maxmium (FWHM) of the PSF peaks at around $1.4-1.5$ arcsec, much larger than those of typical weak lensing surveys ($< 1$ arcsec). Nevertheless, it is a very rich source of imaging data due to its width (covering about $10000 \deg^2$) and depth (redshift distribution peaks at $0.5-0.6$). We therefore have a strong motivation to apply the FQ method on such a data set, not only for testing the shear recovery accuracy on poorly resolved images, but also for generating so far the largest scientifically useful shear catalog.

In \S\ref{FQ}, we briefly review the FQ method, and give an explanation for its robustness at the "small-size" end. In \S\ref{proc}, we outline our main steps in processing the DECaLS images. In particular, we present a novel and efficient way of defining the point sources for PSF reconstruction, and a way of constructing super-flat fields for the DECaLS data. We quantify the multiplicative and additive biases of our shear catalog using the field distortion signals that are available from the data itself.  Finally, in \S\ref{conclusion}, we give our conclusions, and discuss about issues that can affect the quality of our shear catalog.  

\section{The Fourier\_Quad Method} 
\label{FQ}

The FQ method is a particularly simple method. Its shear estimators are made of the multipole moments of the 2D galaxy power spectrum, which are defined as:
\begin{eqnarray}
\label{estimator}
G_1&=&-\frac{1}{2}\int d^2\vec{k}(k_x^2-k_y^2)T(\vec{k})M(\vec{k})\\ \nonumber 
G_2&=&-\int d^2\vec{k}k_x k_y T(\vec{k})M(\vec{k})\\ \nonumber
N&=&\int d^2 \vec{k}\left[k^2-\frac{\beta^2}{2}k^4\right]T(\vec{k})M(\vec{k})
\end{eqnarray}
In the above formulae, $M(\vec{k})$ is the galaxy power spectrum subtracted by terms related to the background noise and the Poisson noise according to eq.(4.9) of \cite{Zhang2015}. $T(\vec{k})$ is the factor which converts the PSF to a Gaussian form, i.e.:   
\begin{equation}
T(\vec{k})={\big|\widetilde{W}_{\beta}(\vec{k})\big|}^2/{\big|\widetilde{W}_{PSF}(\vec{k})\big|}^2\label{tk}.
\end{equation}
Here $\widetilde{W}_{PSF}(\vec{k})$ is the power of the PSF, and $\widetilde{W}_{\beta}(\vec{k})$ $\left[=\exp(-\beta^2\big|\vec{k}\big|^2/2)\right]$ is the counterpart of a Gaussian kernel with the scale radius $\beta$. The choice of $\beta$ should be somewhat larger than the scale radius of the PSF to avoid numerical instability in the conversion. Without any assumptions on the galaxy or PSF morphology, one can show rigorously that each (reduced) shear component can be recovered by taking the ratio between the ensemble averages of the corresponding shear estimators, i.e.:
\begin{equation}
\label{ave}
\frac{\langle G_i\rangle}{\langle N\rangle}=g_i+O(g^3_{1,2})\quad \quad (i=1,2).
\end{equation}

We want to stress here that mathematically, eq.\ref{ave} is well defined for sources of arbitrarily small sizes, albeit with increased statistical noise. In the limit of minuscule source size, the part of $G_i$ and $N$ carrying the source shape information diminishes, while the amplitude of the part contributed by noise stays constant. Note that the form of statistics defined in eq.\ref{ave} is different from the conventional weighting scheme ($\langle w\cdot e_i\rangle/\langle w \rangle$) because the conventional weights are never negative, while the quantity $N$ in the denominator of eq.\ref{ave} can take either positive or negative values. It is exactly this feature that allows the FQ method to correctly present the shear estimators of tiny sources in the form of zero-divide-zero, so that they can be safely included in the shear statistics without incuring systematic errors.    

More recently, as described in \cite{Zhang2016}, the shear statistics can be taken in yet another way (called PDF-SYM) in FQ: by searching for the shear components $(g_1, g_2)$ that can best symmetrize the probability distribution functions (PDF) of the shear estimators $\left[G_1-g_1(N+U), G_2-g_2(N-U)\right]$ around zero. The additional term $U$ is defined as:
\begin{equation}
\label{estimatorU}
U=-\frac{\beta^2}{2}\int d^2 \vec{k}\left(k_x^4-6k_x^2k_y^2+k_y^4\right)T(\vec{k})M(\vec{k}),
\end{equation}
which is a spin-4 quantity introduced for properly taking into account the parity property. It turns out that the PDF-SYM algorithm automatically leads to the minimum statistical error (the Cramer-Rao Bound) in shear recovery, without the need to specify any sort of galaxy weightings. Interestingly, the new algorithm inherits the advantage of the averaging method (eq.\ref{ave}) in handling small sources. For example, point sources only contribute zeros (in the absence of noise) or random numbers centered around zero (in the presence of noise) to the ensemble of $G_{1,2}$, $N$, and $U$, their presence therefore does not affect the symmetry of the PDF no matter what the value of $g_1$ or $g_2$ is. 

Based on the above facts, we are confident to apply the FQ method on images of very general observing conditions. Our shear statistics are all carried out using the PDF-SYM algorithm throughout this work.

\section{Processing of DECaLS Images} 
\label{proc}

We obtained our DECaLS data from the website of the DESI Legacy Imaging Surveys\footnote{www.legacysurvey.org} in three bands: g, r, and z, which contain 21925/22592/22941 exposures respectively. The image file names all end with "ooi", meaning that they have been proccessed through the "Community Pipeline", and their sky backgrounds are not subtracted. These images were taken by the Dark Energy Camera (DECam) on the Blanco 4m telescope of the Cerro Tololo Inter-American Observatory in the year between 2013 and 2019. 

\begin{figure}
	\centering \includegraphics[scale=0.5]{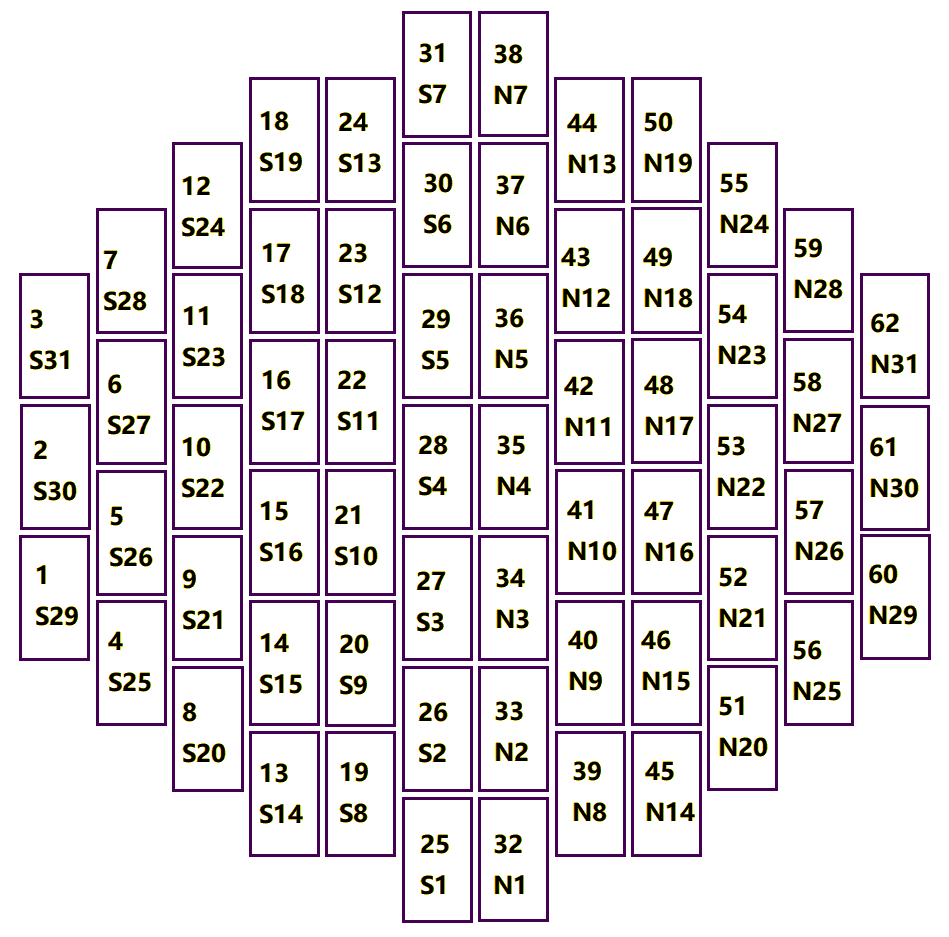}
	\caption{The distribution of the 62 CCD chips on the focal plane. Inside each chip, we show the CCD number on the upper side, and the detector position ID on the lower side.}
	\label{ccd_show}
\end{figure}

Each DECaLS exposure contains 62 chips, the distribution of which is shown in fig.\ref{ccd_show}. We use the CCD number shown in the figure to label each CCD in the paper. Each exposure is independently processed by our FQ pipeline, which is previously used for processing the CFHTLenS data. It contains all the neccessary steps for shear measurement \citep{Zhang2019}, including: background removal, identification of cosmic ray and other image defects, astrometric calibration, source/noise identification, PSF reconstruction, calculation of the shear estimators in Fourier space. Most of the above steps are carried out independently on each chip, except for astrometric calibration and PSF reconstruction. Our pipeline does not contain the part for photometric calibration. We use the source catalog (with photo-z) from \cite{Zou2019} to locate the galaxies. After removing problematic exposures, our final shear catalog contains 15420/15162/16501 exposures for the g/r/z band respectively. Using the photo-z catalog, we include in our pipeline an additional step to remove the overlapped/blended sources with redshift difference $\Delta z$ larger than 0.1. For those with $\Delta z \le 0.1$, we simply treat them as a single source. Note that in the later case, the resulting irregular shape of the source is not a trouble for FQ at all.

\begin{figure}
	\centering \includegraphics[scale=0.4]{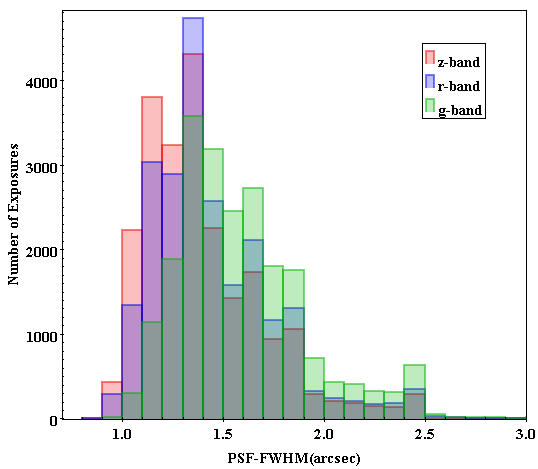}
	\caption{The distribution of the PSF FWHM for the g, r, z bands.}
	\label{psf_size}
\end{figure}

Overall, the version of our FQ pipeline for processing the DECaLS data is very similar to the one described in \cite{Zhang2019} (Z19 hereafter). The main difference is about the PSF size, which is much larger on average in DECaLS. Fig.\ref{psf_size} shows the distribution of the PSF FWHM for the g, r, z bands in the green, blue, and red colors respectively. All three distributions peak at around $1.4$ arcsec, which is much larger than that of the CFHTLenS ($\sim 0.6 - 0.8$ arcsec). Because of the large PSF size, we need to remove the "noise reduction" step in the FQ pipeline to avoid shear bias. In the rest of this section, we give some details regarding the other main adjustments of our pipeline.

\subsection{PSF Reconstruction}
\label{PSF_recon}

\begin{figure*}
	\centering
	\includegraphics[width=1.0\textwidth]{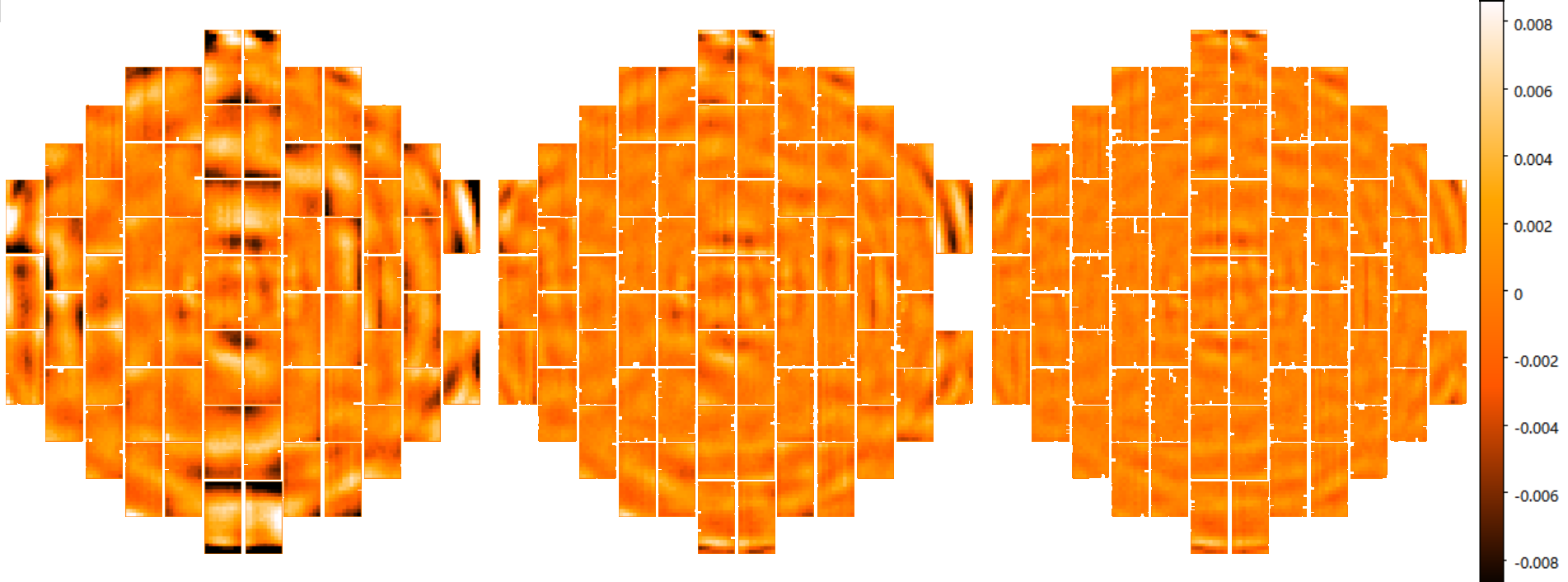}
	\caption{The residual of the $e_1$ component of the PSF ellipticity from polynomial fitting up to the first (left plot), second (middle plot), and third order (right). The data is from the z band.}
	\label{PSFe1}
\end{figure*}
\begin{figure*}
	\centering
	\includegraphics[width=1.0\textwidth]{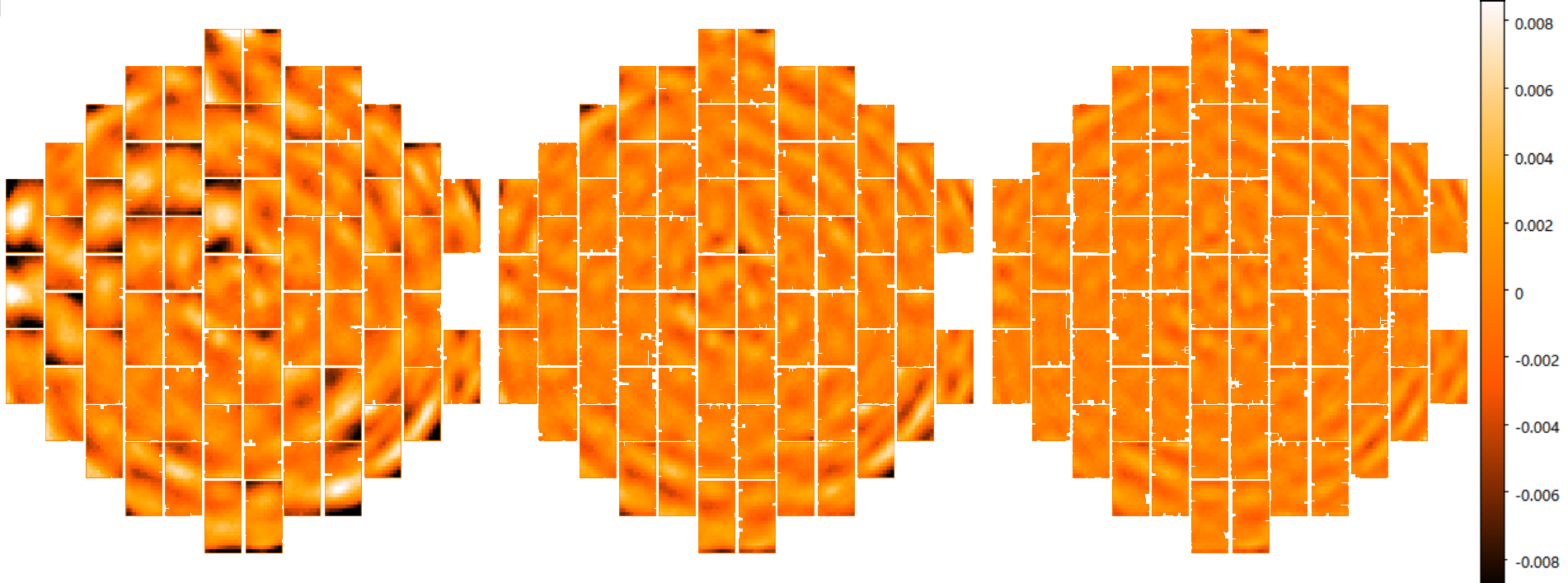}
	\caption{Same as fig.\ref{PSFe1}, but for the $e_2$ component. }
	\label{PSFe2}
\end{figure*}
Given that the pixel size of DECam is about 0.263 arcsec, and the PSF FWHM can reach more than 2 arcsec, we set our image stamps to be $64\times 64$ so that they are large enough to properly contain the majority of sources. Selecting out stars for PSF reconstruction can be very challenging in the case of poor seeing. We find that the algorithm described in Z19 is not stable enough for our purpose here. Our new method turns out to be more accurate and stable, and completely automatic. It makes use of the fact that {\it the PSF stars should have profiles that are similar to their neighbors on the same chip}. To do so, we propose a way to quantify the similarity between any two images (in Fourier space). The friends-of-friends (FOF) method can then be used to identify the largest group of sources in each chip that are similar enough to each other. These sources are defined as point sources/stars for PSF reconstruction. More specifically, it includes the following steps:

1. We select out all the sources with SNR$\ge 100$ as the candidates of the PSF stars. If the whole exposure contains less than about 200 such sources, we do not continue (not enough candidates), and mark the exposure as a bad one. 

2. For each good exposure, we take the 2D power spectrum $P(\vec{k})$ of each candidate source, and measure its area, which is defined as the number of pixels above 2\% of $P(\vec{k}=0)$. We then sort the candidates according to their areas, and take the top 1/3 as the reference sample, as they are very likely to be true stars. Note that in Fourier space, point sources should have the most extended profiles. 

3. We define the morphological distance $D_{nm}$ between two power spectrum images (with indices n and m respectively) as:
\begin{equation}
D_{nm}^2=2\sum_{i=1}^{N}(I_i^n-I_i^m)^2/(I_i^n+I_i^m)
\label{DD}
\end{equation}
where $I_i^n$ refers to the value of the $i^{th}$ pixel on the $n^{th}$ image (normalized). $N$ is the total number of pixels included in the definition. Since most of the morphological information are concentrated in the central part of the Fourier space, we only include the central $32\times 32$ of the power spectrum to calculate $D_{nm}$. 

4. Within the reference sample on each chip, we calculate the mutual distances between the candidates. The results from different chips of the exposure are then combined to form an ensemble, from the PDF of which we can identify a significant peak at $\bar{D}$ at the lower end with a scattering radius $\sigma$.  
   
5. We use $\bar{D}+4\sigma$ as the threshold to define the FOF groups within each chip. The group that contains the largest number of members is believed to contain the true point sources. The chosen group is required to contain more than 16 members, otherwise the chip is not processed further.

\begin{figure*}
	\centering
	\includegraphics[width=1.0\textwidth]{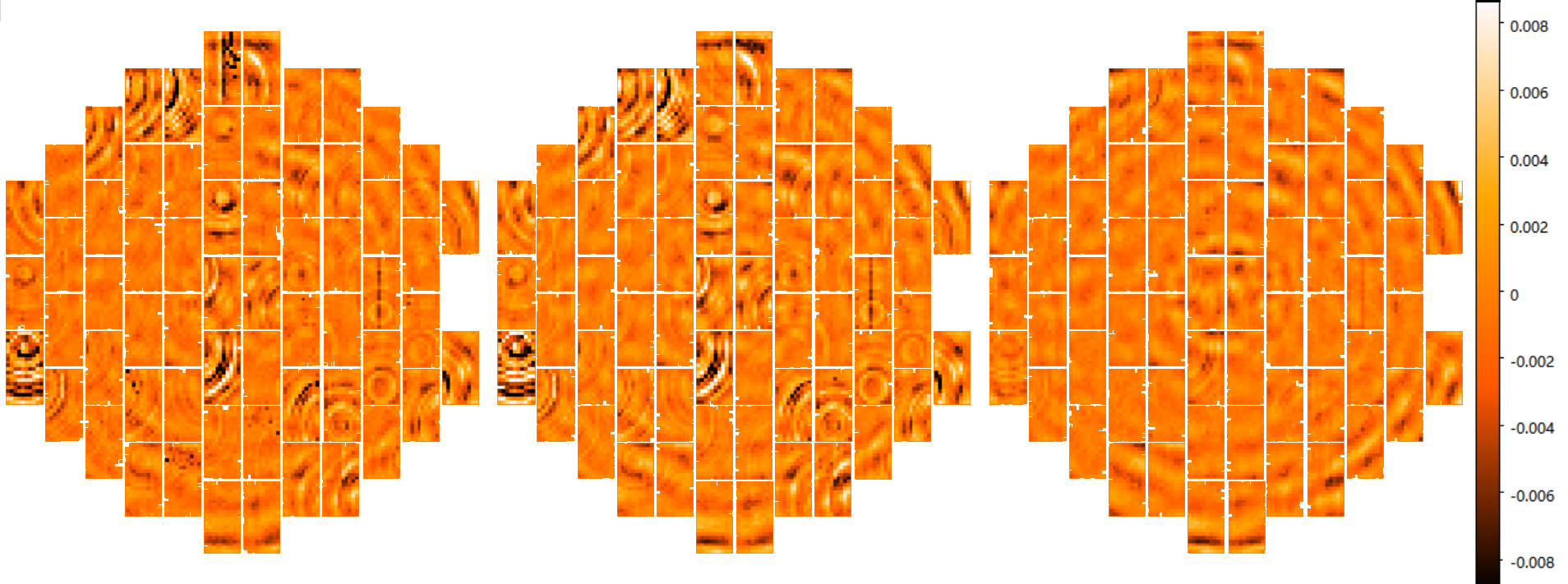}
	\caption{The residual of the PSF area from polynomial fitting up to the third order. From left to the right are the results for the g, r, and z band respectively. }
	\label{PSFsize_grz}
\end{figure*}

After collecting the stellar images, we use the polynomial functions as the templates to model the spatial variations of the PSF morphology within each chip. It is worth noting that in FQ, we need to reconstruct the power spectrum of the PSF, rather than its real space image. This fact makes it extremely convenient in image alignment. The PSF power spectrum $W_{PSF}$ centered at the position of $(x,y)$ on the chip is parameterized as:
\begin{equation}
\label{PSF_func}
W_{PSF}(x,y)=\sum_{i=0}^n\sum_{j=0}^iA_{ij}x^jy^{i-j}.
\end{equation}
In the definition, $n$ is the maximal order of the polynomial functions in the fitting, and each $A_{ij}$ is a $64\times 64$ image derived from the pixel-by-pixel fitting to the stellar power spectra. 

In fig.\ref{PSFe1} and fig.\ref{PSFe2}, we show the residual of the PSF ellipticity components $e_1$ and $e_2$ after fitting with polynomial functions up to the 1st, 2nd, and 3rd order in z-band. The ellipticity components are defined in the conventional way as $e_1=(Q_{20}-Q_{02})/(Q_{20}+Q_{02})$, $e_2=2Q_{11}/(Q_{20}+Q_{02})$, except that the moments $Q_{ij}$ are defined in Fourier space as: 
\begin{equation}
Q_{ij}=\sum_{P(k)>P_0}P(k)k_x^ik_y^j.
\end{equation}  
The threshold $P_0$ is again chosen to be $0.02P(\vec{k}=0)$.  According to the figures, it is clear that the first and second order polynomial fitting does not seem accurate enough. This conclusion is different from what we got earlier using the CFHTLenS data \citep{Lu2017}, for which the first order chipwise polynomial fitting is found to be good enough. Our final shear catalog is therefore based on the third order polynomial fitting functions. Further increasing the order number would result in somewhat minor improvement, but would require more stars per chip to avoid overfitting problems.

In fig.\ref{PSFsize_grz}, we show the distribution of the residual (relative difference) of the PSF area that we defined earlier in this section. We only include the results for $n=3$. The three plots from the left to the right are for the g, r, and z band respectively. For the g and r bands, one can see very clearly the features caused by the tree rings on the CCDs \citep{Plazas2014,Jarvis2021}. As we show later in the paper, the quality of shear recovery in these two bands is not as good as in z-band.

\subsection{Super-Flat Fields}
\label{superflat}

The images we obtained had been previously corrected by flat fielding to a certain level. Such correction is essential for the shear measurement because the inhomogeneous responsivities of the CCD pixels can directly affect the galaxy shapes. For the purpose of this work, we care most about the flat field accuracy on the scales comparable to typical galaxy sizes. Study of this issue requires us to reconstruct some kind of super-flat field, i.e., a flat field with higher accuracy. Fortunately, there is a way to do so given that we have more than twenty thousand exposures available in each band. 

Our idea is to stack the CCD images normalized by their background fields. The sky background in each CCD is modelled with low order 2D polynomial functions, therefore it is quite smooth on small scales. For this reason, we believe that the small scale features of the flat field should show up after stacking many CCDs on the same position of the focal plane. In doing so, we mask out the pixels that are covered by sources or defects, so that they are not counted in the stacking. In notation, our definition of the super-flat map SF($\vec{x}$) can be described as:
\begin{equation}
SF(\vec{x})=\frac{\sum_{j=1}^N\left[f_j(\vec{x})/f_j^b(\vec{x})-1\right]W_j(\vec{x})}{\sum_{j=1}^NW_j(\vec{x})},
\label{SFlat}
\end{equation}
in which j is the index of the exposure, $f_j$ and $f_j^b$ are the pixel readout and the background respectively, and $W_j$ is the weight with two possible values: one if the pixel is normal, or zero if it is masked. N is the total number of exposures.

The typical amplitudes of the pixels on the super-flat maps are in the range of $10^{-5}-10^{-3}$, rarely reach $10^{-2}$ or above. Overall, we find that the g and r band super-flats have more structures on small scales than those of the z-band.
\begin{figure}
	\centering
	\includegraphics[width=0.45\textwidth]{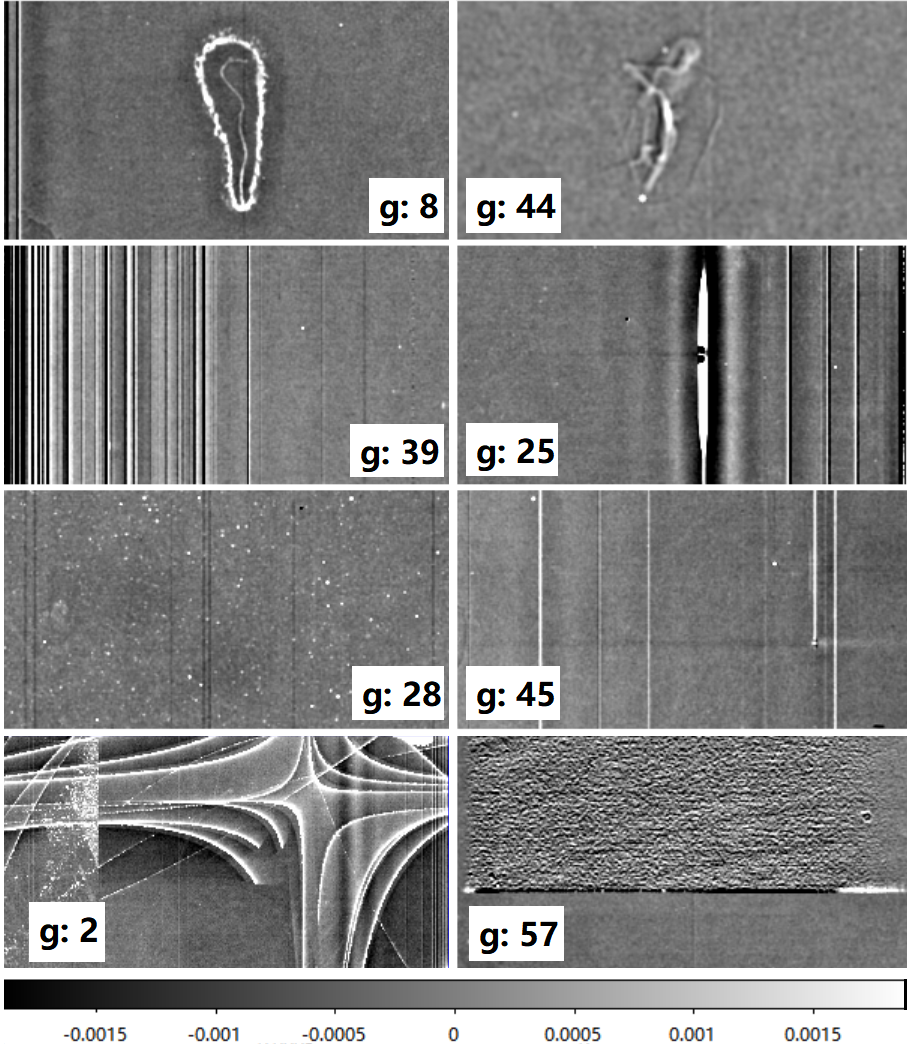}
	\caption{Each panel shows the map of the super-flat on a certain part of a CCD, the number of which is given in the corners with the band name. }
	\label{superflat_g}
\end{figure}
In fig.\ref{superflat_g}, we show some of the features for the g band. In each panel of the figure, there are characters in corner indicating the band name and the CCD number. The problems are summarized below:

1. In the panels of the first row, there appear to be threads or something similar attached to the CCD surface. These features also show up at the same locations of the r and z band data. The problem on the $44^{th}$ CCD is similar, and known to exist from November of 2018 according to the offical website of the Cerro Tololo Inter American Observatory\footnote{https://noirlab.edu/science/programs/ctio/instruments/Dark-Energy-Camera/Status-DECam-CCDs}.

\begin{figure}
	\centering
	\includegraphics[width=0.45\textwidth]{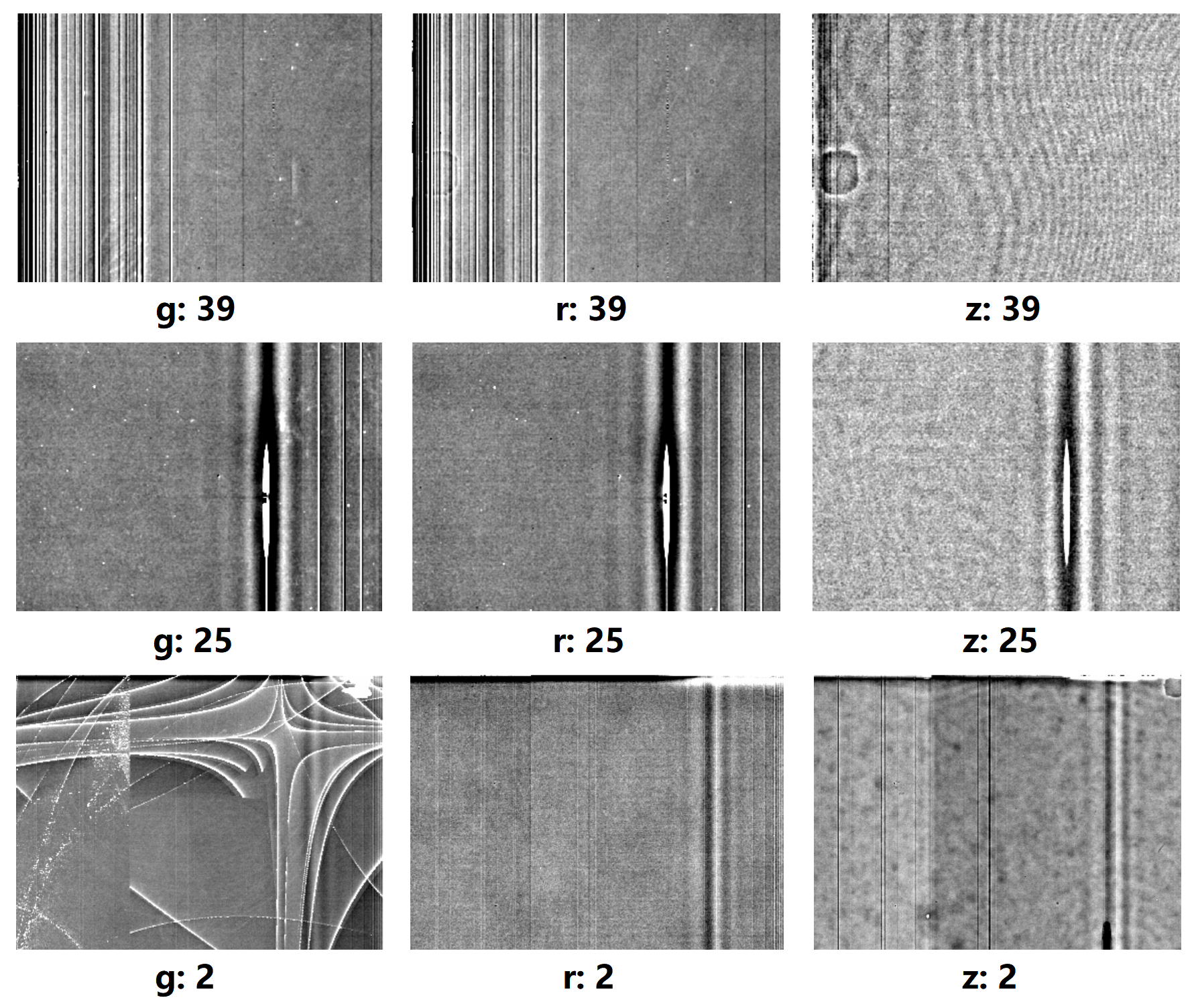}
	\caption{Comparison of the super-flats in the three bands for the $39^{th}$, the $25^{th}$, and the $2^{nd}$ CCDs.}
	\label{superflat_grz_39_25}
\end{figure}

2. On the second row, we can see some significant amount of bad columns parallel to the long side of the CCD. This type of features often show up near the edges of the CCDs, and can be more than 200-pixel wide sometimes. This problem is also listed on the official site of the observatory\footnote{https://noirlab.edu/science/programs/ctio/instruments/Dark-Energy-Camera/Known-Problems}, and called funky columns. The good news is that this problem is sometimes much milder in the z-band. The comparison between the three bands for the $39^{th}$ and $25^{th}$ CCDs are shown in fig.\ref{superflat_grz_39_25}. On the other hand, one can see that the tape bumps as well as the fringe patterns on the CCDs becomes more obvious in z-band. 

\begin{figure}
	\centering
	\includegraphics[width=0.45\textwidth]{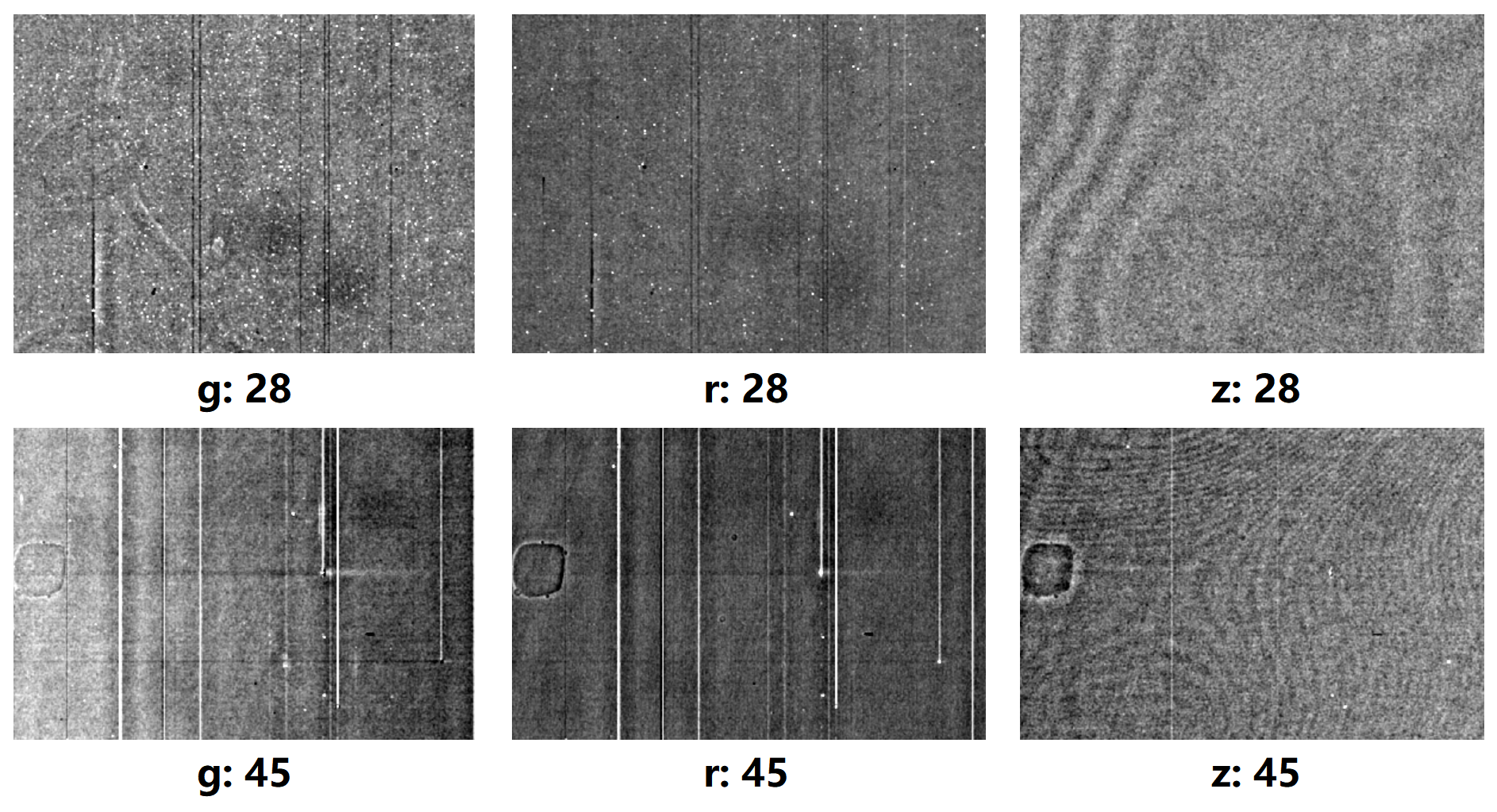}
	\caption{Comparison of the super-flats in the three bands for the $28^{th}$ and $45^{th}$ CCDs.}
	\label{superflat_grz_28_45}
\end{figure}

3. On the third row of fig.\ref{superflat_g}, we show the defects found in the central parts of the $28^{th}$ and $45^{th}$ CCDs: there are many hot spots on the $28^{th}$ CCD, and a number of hot/cold columns on the $45^{th}$ one. Similar problems exist on the r band data, but at a somewhat milder level. Again, the z-band does not seem to be affected by these problems at all. The comparison for the same area of the CCD is provided for the three bands in fig.\ref{superflat_grz_28_45}. 

4. On the bottom of fig.\ref{superflat_g}, we show the problems found on the $2^{nd}$ and $57^{th}$ CCD. The features on the $2^{nd}$ CCD look like scratches on the CCD surface. This type of phenomenon also occurs sporadically on some other CCDs in g-band, but never on those of the r or z band. The multi-band comparison for this problem is shown in the botton row of fig.\ref{superflat_grz_39_25}. The $57^{th}$ CCD appears to have a large number of bad pixels on the upper-left corner of the chip. These problems are present at similar levels for all three bands.  

Finally, though not shown here, we note that the $31^{st}$ CCD is known to have the unstable-gain issue. It affects the superflats of all three bands significantly (at a few percent level). 

\begin{figure}
	\centering
	\includegraphics[width=0.45\textwidth]{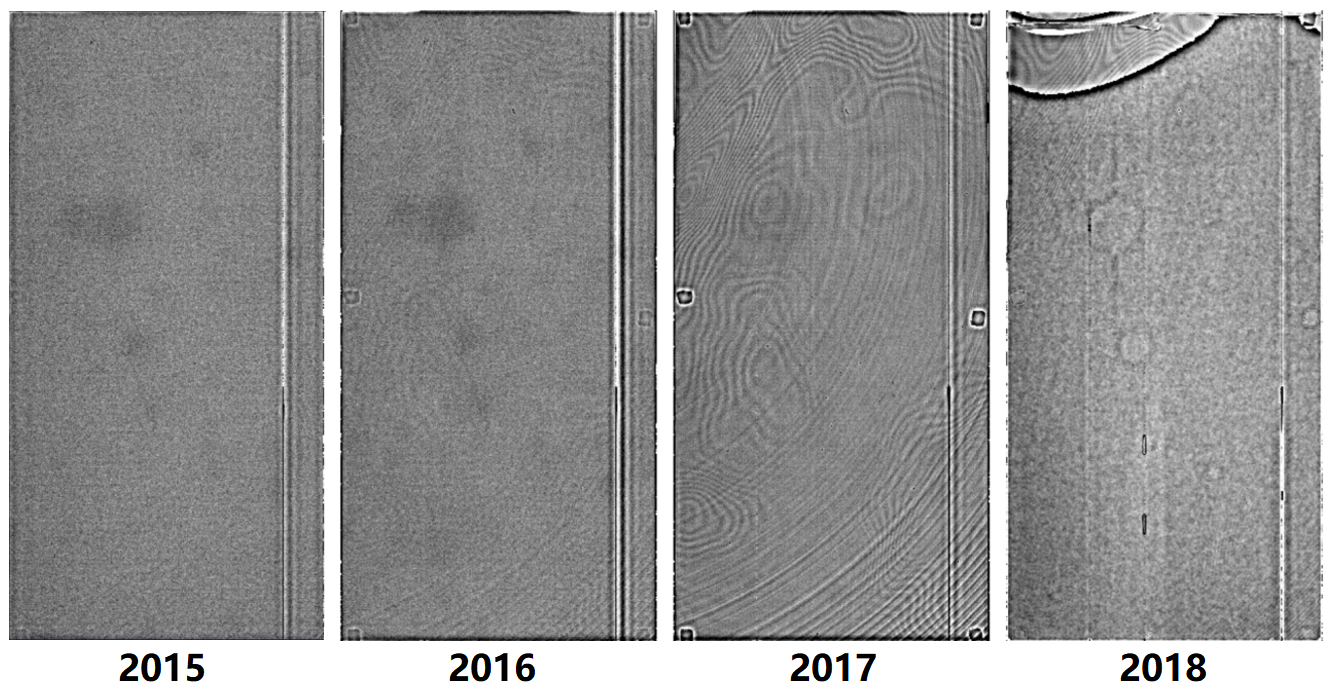}
	\caption{Comparison of the super-flats in four years (2015 - 2018) for the $16^{th}$ CCD of the z-band.}
	\label{superflat_z16_15-18}
\end{figure}
\begin{figure}
	\centering
	\includegraphics[width=0.45\textwidth]{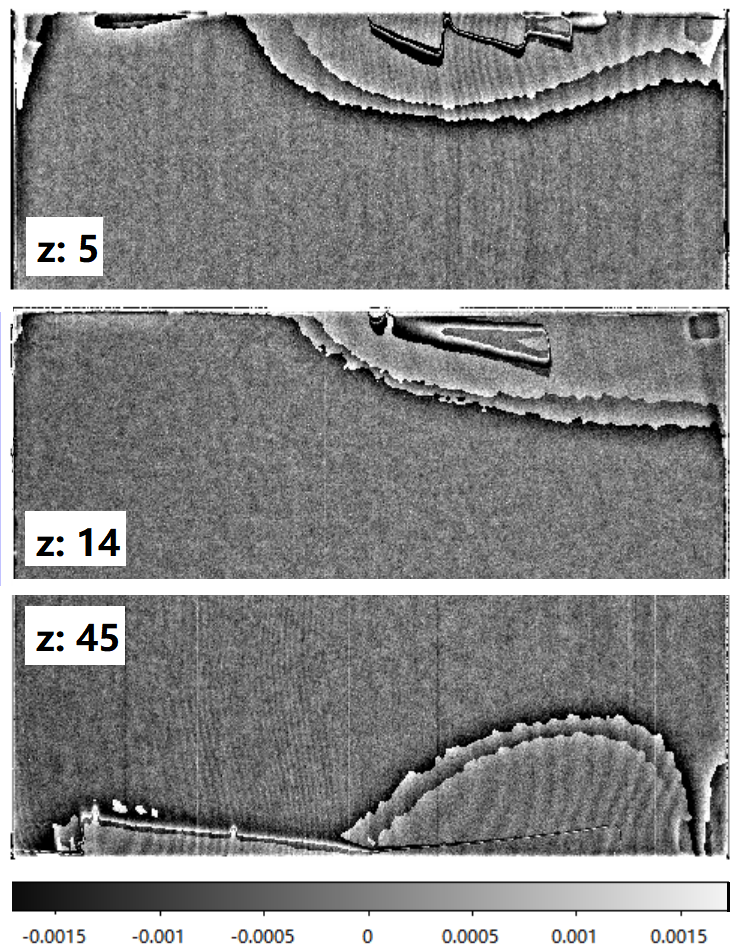}
	\caption{Defects revealed by the super-flats of year 2018 on the $5^{th}$, $14^{th}$ and $45^{th}$ CCD in z-band.}
	\label{cracks}
\end{figure}

The above problems are discovered by stacking all the available exposures in each band, which were taken over many years. To find out the time dependence of the super-flats, we simply stack the ones within one year for four consecutive years (2015 - 2018), as shown in fig.\ref{superflat_z16_15-18} for the $16^{th}$ CCD of the z-band. There are indeed some visible changes between the results of different years, especially for the year of 2017 and 2018. The 2018 result even reveals cracks or something similar on multiple layers. This phenomenon occurs on a number of CCDs, all starting from 2018, as shown in fig.\ref{cracks}. These changes could be caused by earthquakes near the location of the telescope.
 
The super-flats reveal some intrinsic problems of the CCDs, which appear to be dependent on the photon wavelength and time. To take these facts into account for the shear measurement, we adopt the following simple recipe: we divide each CCD pixel value at position $\vec{x}$ by $1+SF(\vec{x})$ if $\vert SF(\vec{x})\vert < 0.02$, or otherwise mark it as a bad pixel. To account for the time dependence of the super-flats, for each band, we group the exposures according to their observing time, so that each group contain at least a few hundred exposures, usually covering a few months in observation time. The super-flat field is created and used within each group.

We are aware that our way of correcting the pixel values may not be correct. However, the main purpose of our study in this part is about finding out how important these effects are. Fortunately, as we show in the next section, the quality of our shear catalog is not significantly affected by the above procedures.

\subsection{Field Distortion Test}
\label{fd}

\begin{figure*}
	\centering
	\includegraphics[width=1.0\textwidth]{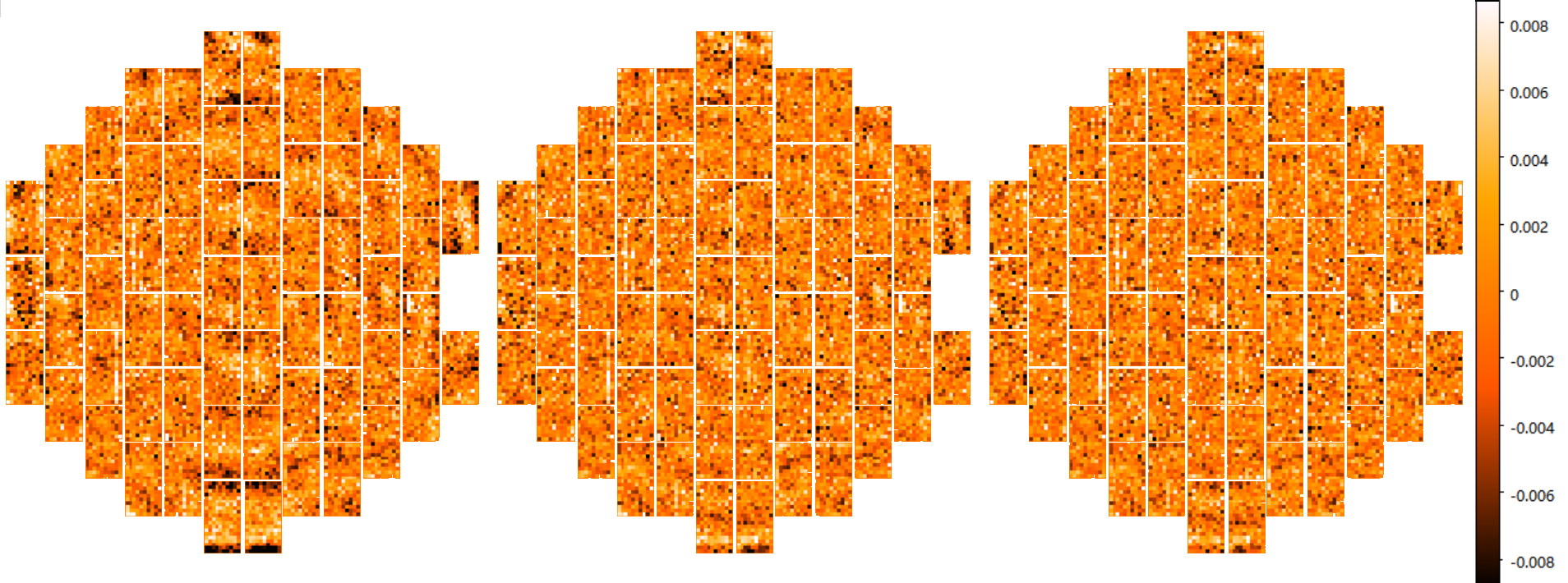}
	\caption{The distribution of the average $g_1$ shear component for PSF polynomial fitting up to the first (left plot), second (middle plot), and third order (right). The data is from z-band. }
	\label{g1}
\end{figure*}
\begin{figure*}
	\centering
	\includegraphics[width=1.0\textwidth]{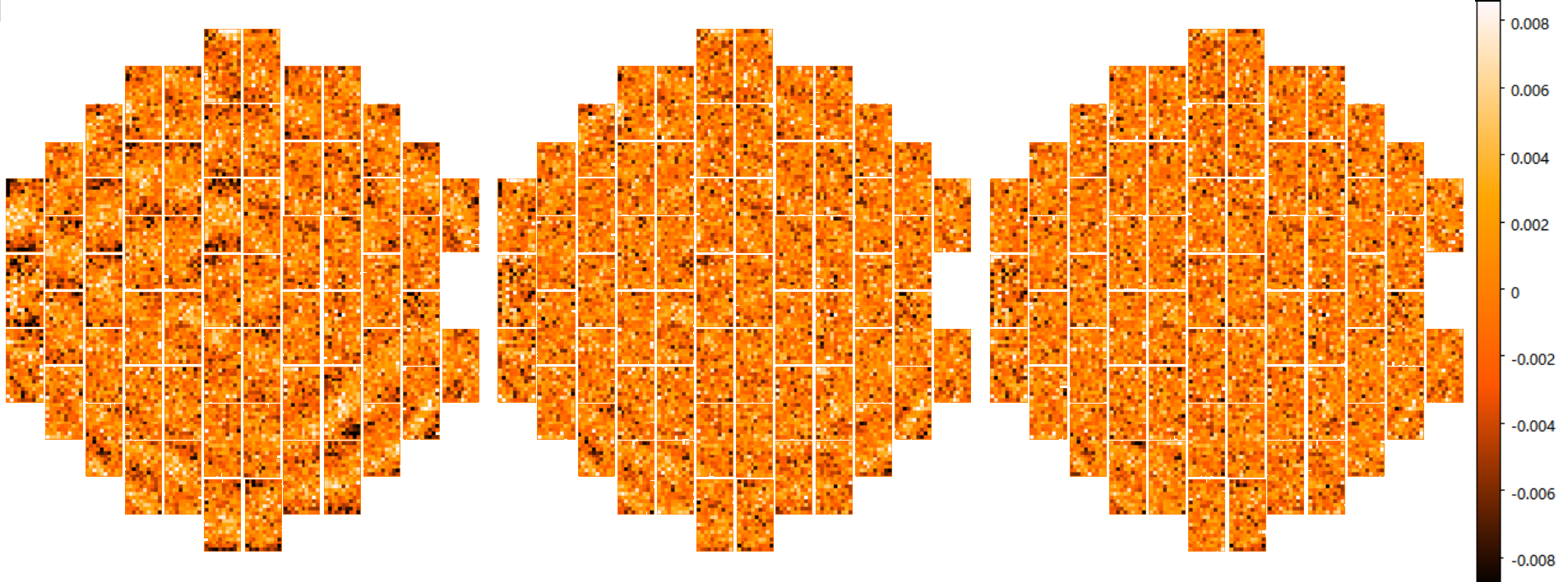}
	\caption{Same as fig.\ref{g1}, but for the $g_2$ component. }
	\label{g2}
\end{figure*}
As pointed out by Z19, a direct way of testing the quality of the shear catalog is to compare the galaxy shear with the field distortion (FD hereafter), which is the spatial deformation of the source images caused by optical aberration. The FD signals can be recovered using the parameters from the astrometrical calibration. Its amplitude is about a few times $10^{-3}$ for CFHTLenS and DECaLS, which is comparable to the cosmic shear signals, therefore ideal for such a test. 

It is worth noting that in the FD test, since we stack the galaxy shear estimators at fixed locations of the CCDs (because the distribution pattern of FD is usually quite stable on the focal plane) to recover particular values of FD, any systematic effect (from hardware, background, astrometry, PSF, etc.) that varies spatially on the focal plane can in principle be captured, no matter whether the effect is known or not known beforehand. In other shear measurement methods, it is currently common to calibrate the shear recovery accuracy with simulations. The simulation test is valuable because of its flexibility in design for studying various known effects, such as image blending, but not for unknown ones at all. In this sense, the FD test can be regarded as an important complement to the simulation test. We will report the result of a comprehensive simulation test for our FQ pipeline in a separate work.    

Before showing the results of the FD test, it is perhaps useful to first take a look at the results of a null test, i.e., the distribution of the stacked shear signals (FD signal removed) on the focal plane. These are shown in fig.\ref{g1} and \ref{g2} for $g_1$ and $g_2$ in the z-band respectively. In each figure, the results are presented for three choices of the polynomial order n in PSF reconstruction. It is again quite obvious that the residuals are obvious for both $g_1$ and $g_2$ when n=1, and the patterns are similar to those shown in the leftmost plots of fig.\ref{PSFe1} and \ref{PSFe2}. It demonstrates the well known fact that PSF error is among the most important reasons for shear bias. For n=2 or 3, it is not so easy to judge the quality of the shear catalog from the plots because of the noise.

\begin{figure}
	\centering
	\includegraphics[width=0.45\textwidth]{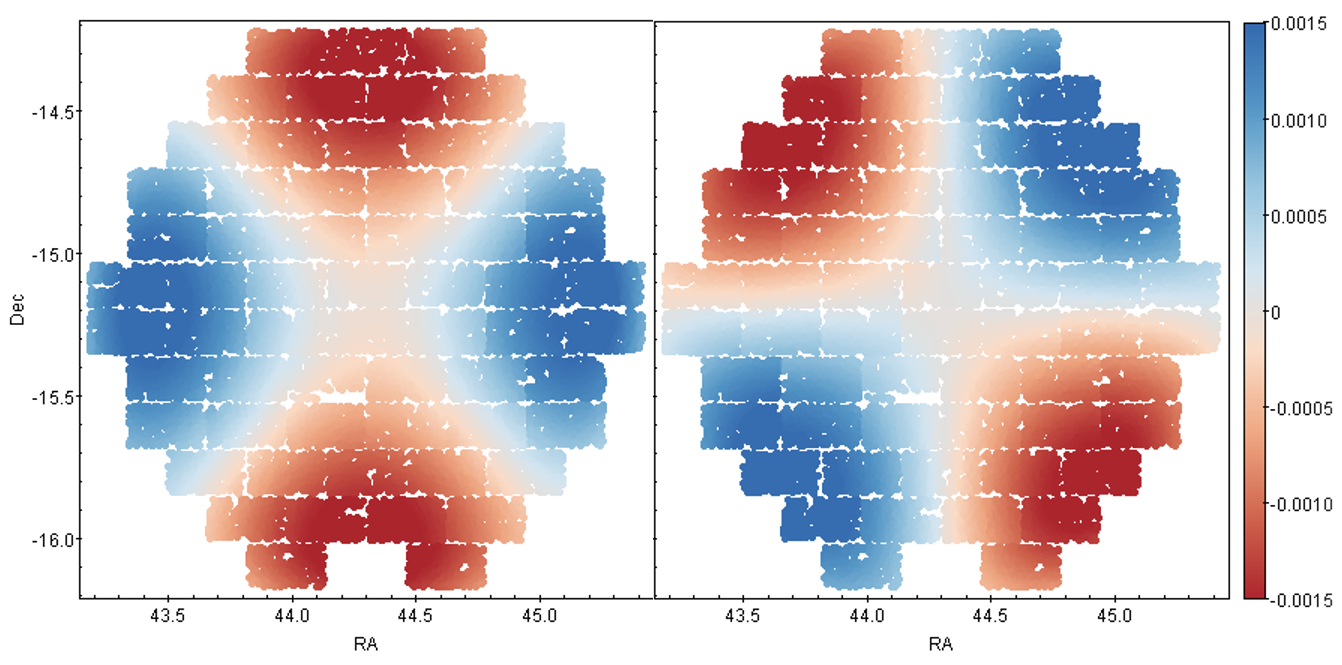}
	\caption{The distribution of the field distortion signal on the CCDs. The left and right plots show the patterns of $g^f_1$ and $g^f_2$ respectively. The components of $g^f$ are defined in the coordinates of RA and Dec.}
	\label{FD}
\end{figure}
The FD test can be regarded as a way of increasing the signal-to-noise ratio of the null test on the focal plane. It is also better than a global null test in the sense that the FD test can provide estimates of both the multiplicative and additive shear biases, while the global test only gives the additive biases. In fig.\ref{FD}, we show the distribution of the field distortion signals ($g^f_1$, $g^f_2$) on the focal plane. The amplitudes of the FD signals are less than about 0.0015, much smaller than the typical values of CFHTLenS ($\sim 0.005$).

\begin{figure}
	\centering
	\includegraphics[width=0.45\textwidth]{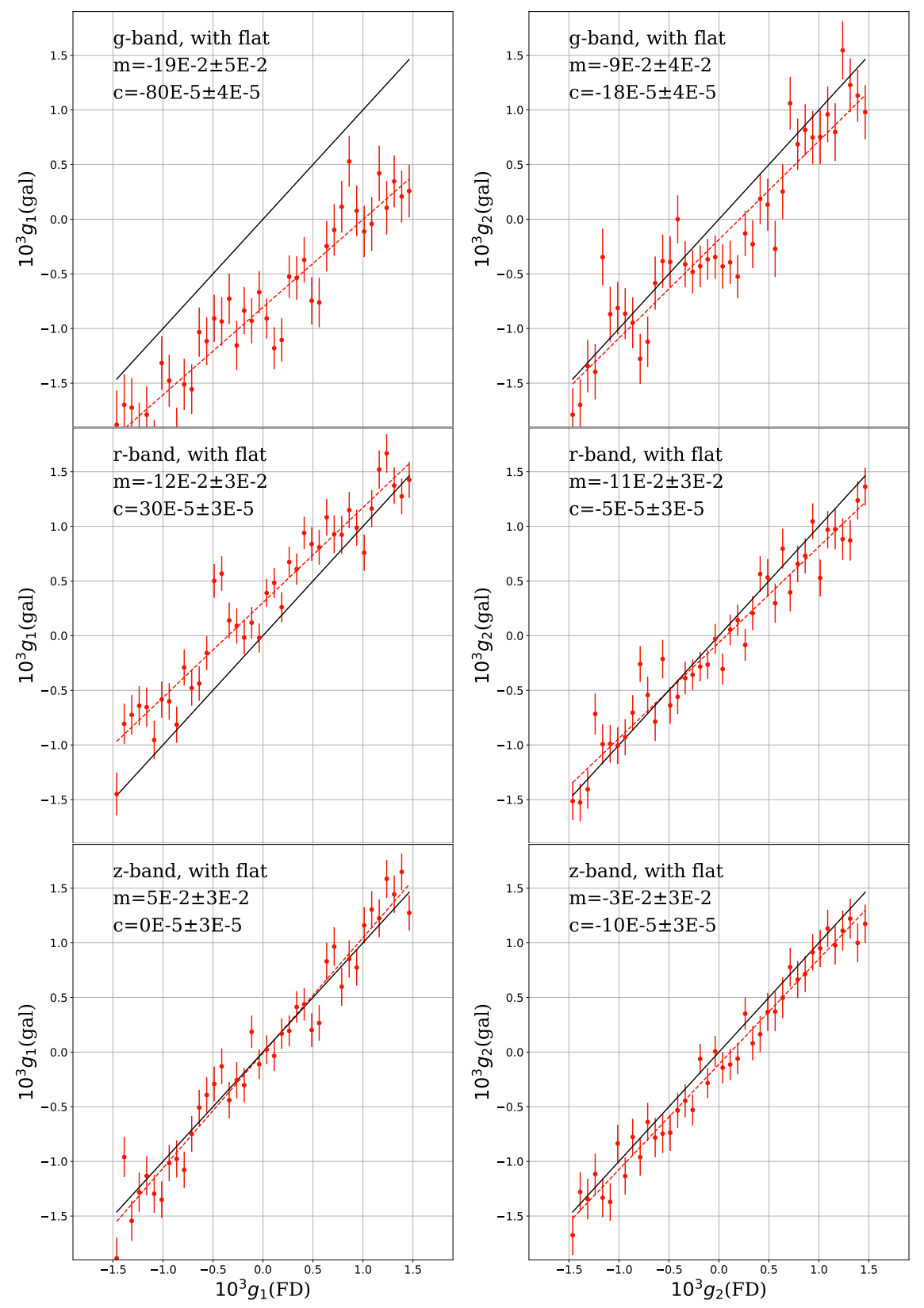}
	\caption{The results of the FD test for galaxy images in the g, r, and z band respectively. In each plot, the black solid line is the "y=x" reference, and the red dashed line is the best fit of the data points. The multiplicative and additive biases are shown in the upper left corner of each plot. The super-flat fields are used in the pipeline.}
	\label{FD_grz_with_flat}
\end{figure}
\begin{figure}
	\centering
	\includegraphics[width=0.45\textwidth]{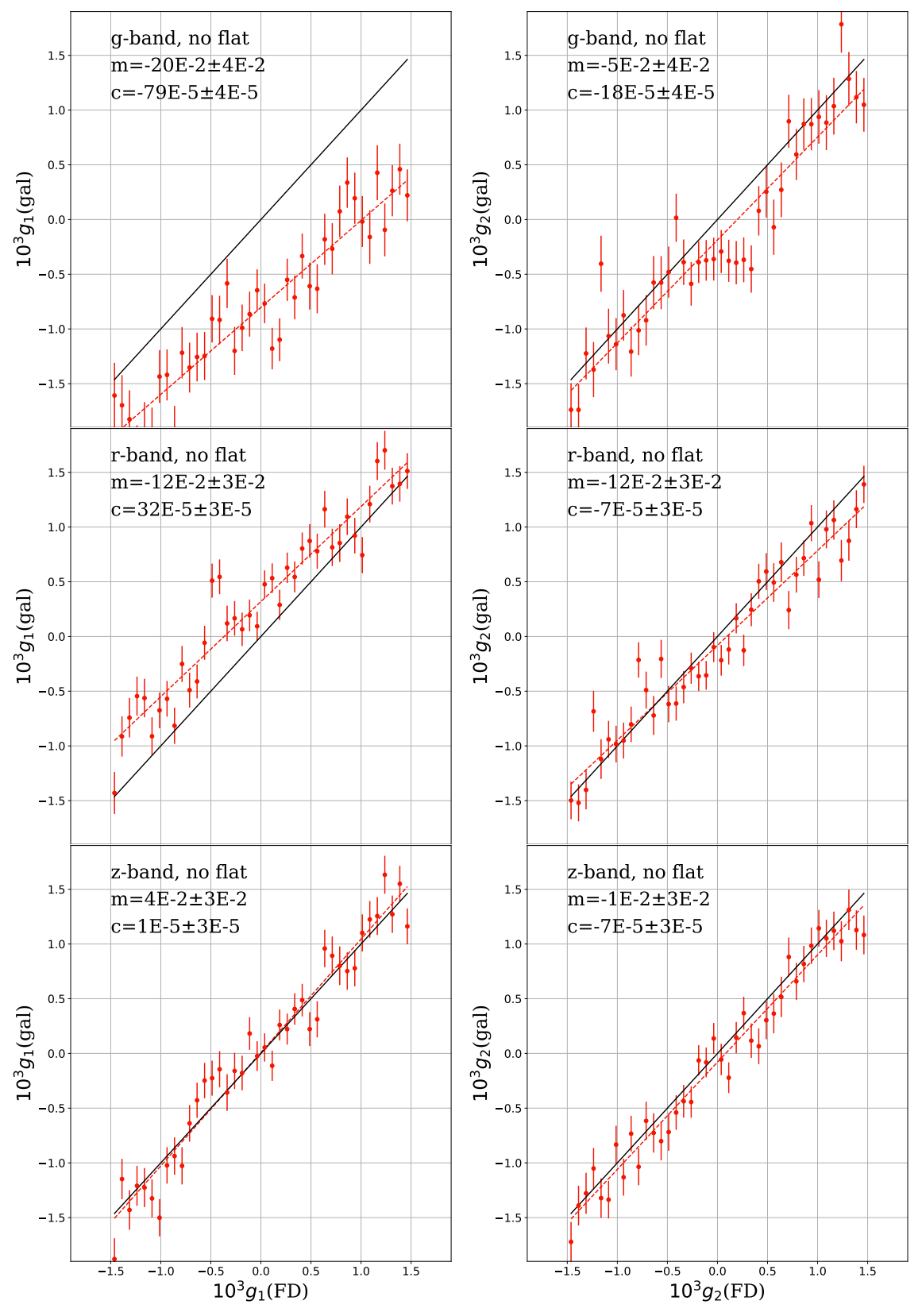}
	\caption{Similar to fig.\ref{FD_grz_with_flat}, except that the shear catalogs are made without using the super-flat fields.}
	\label{FD_grz_no_flat}
\end{figure}

The comparison between the galaxy shear and the FD is shown in fig.\ref{FD_grz_with_flat} and \ref{FD_grz_no_flat} for the cases of including and not including the super-flats in our image processing respectively. As one can see from the figures, the influence of the super-flats is quite marginal. This is perhaps a good news, given that we do not yet have a robust way of treating the effects found in the super-flats. It is worth noting that the results in fig.\ref{FD_grz_with_flat} and \ref{FD_grz_no_flat} do not involve any calibrations. It is therefore very encouraging to see that the quality of the z-band catalog is exceptionally good. The effective shape noise per galaxy ($\sigma_{\epsilon}$) is around $0.28$. The r-band quality is also acceptable, because the multiplicative biases are mainly driven by the data at around $g_1^f=-5\times 10^{-4}$ and $g_2^f=-7\times 10^{-4}$, which can be excluded in scientific applications. Its additive bias in $g_1$ should also be corrected. The average shape noise of the r-band images is very similar to that of the z-band. In these tests, we use the sources with $\nu_F>4$, where $\nu_F$ is the selection function proposed by \cite{Li2020} for avoiding selection effects.

Finally, the figures also show that the g-band catalog is significantly worse than the other two bands. We therefore do not recommend using it, although, if one wishes, the g-band catalog can still be useful after one applies the corresponding multiplicative and additive corrections, as what is usually done. The average shape noise in g-band is 0.31, also somewhat larger than those of the other two bands. 

\begin{figure}
	\centering
	\includegraphics[width=0.45\textwidth]{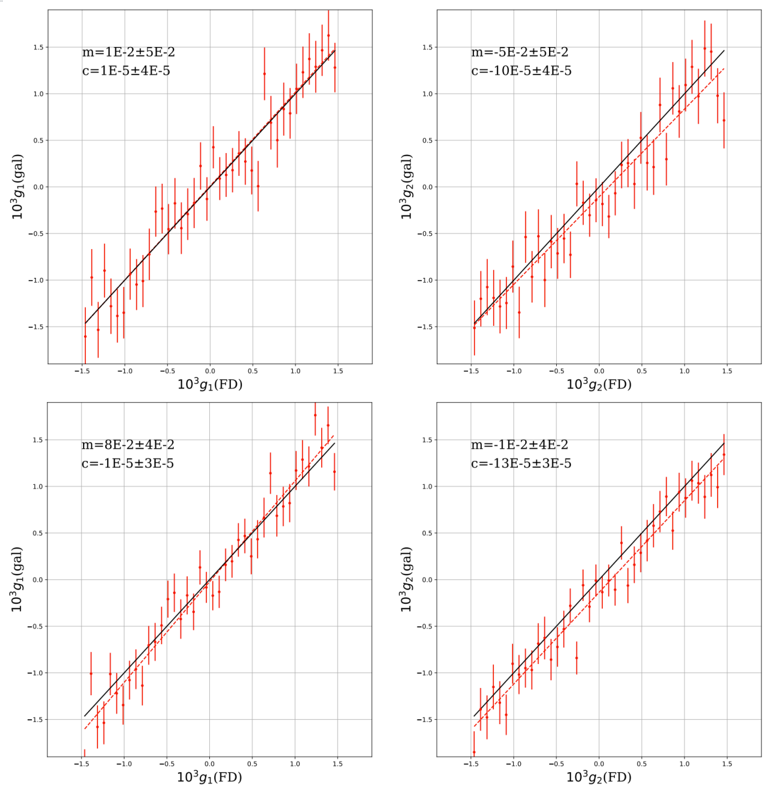}
	\caption{The results of the FD test for galaxies of two catagories: those with PSF FWHM greater (upper panels) or smaller (lower panels) than 1.4 arcsec. In each plot, the black solid line is the "y=x" reference, and the red dashed line is the best fit of the data points. The multiplicative and additive biases are shown in the upper left corner of each plot.}
	\label{FD_test_large_small_PSF}
\end{figure}

In fig.\ref{FD_test_large_small_PSF}, we show the results of the FD test in z-band (with the super-flat correction) for two cases: the upper panels show the comparison between the galaxy shear $g_{1,2}$(gal) and the FD signal $g_{1,2}$(FD) for images with large PSF FWHM ($> 1.4$ arcsec), and the lower panels show the results for the case of small PSF (FWHM $< 1.4$ arcsec). It is very encouraging to note that the PSF size does not affect the quality of shear recovery in our catalog. Indeed, in both cases, a large fraction of sources have sizes that are comparable to the PSF size. Our results therefore demonstrates the robustness of the FQ pipeline in processing poorly resolved images. The two samples are mainly different in their average shape noises: we find that $\sigma_{\epsilon}\approx 0.31$ for the sample with PSF FWHM$> 1.4$ arcsec, and only 0.26 for the other sample.

\subsection{Catalog Properties}
\label{fd}

\begin{figure}
	\centering
	\includegraphics[width=0.45\textwidth]{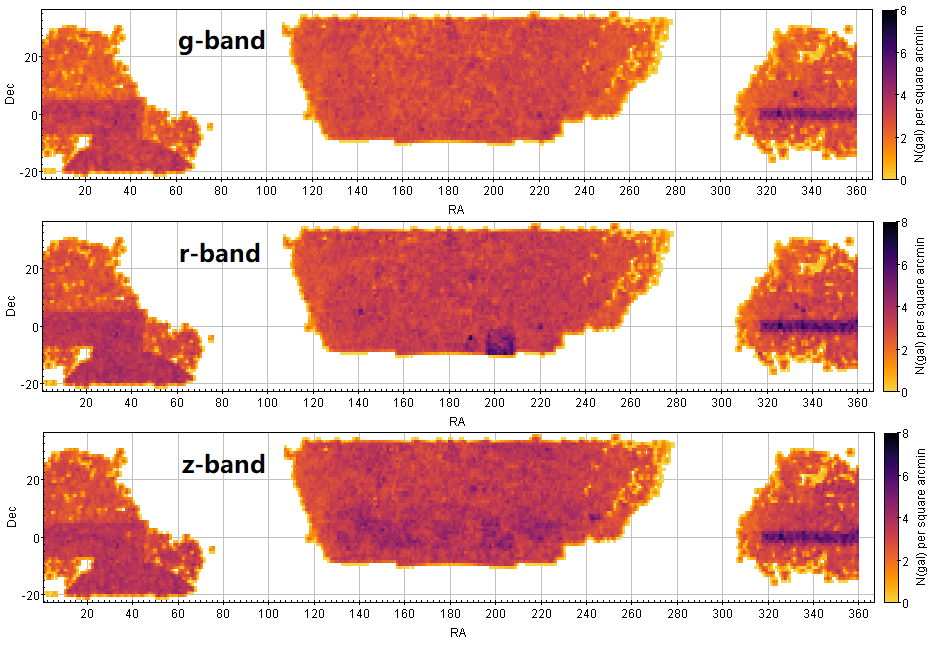}
	\caption{The distribution of the source galaxies in the g, r, and z bands. }
	\label{gal_density}
\end{figure}

\begin{figure}
	\centering
	\includegraphics[width=0.45\textwidth]{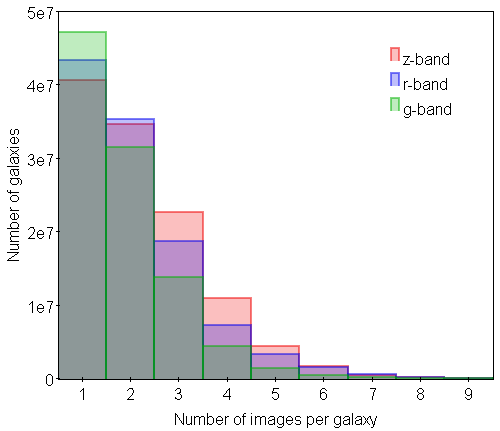}
	\caption{Distribution of the number of images per galaxy in the g, r, and z bands. }
	\label{num_image_per_gal}
\end{figure}

\begin{figure}
	\centering
	\includegraphics[width=0.45\textwidth]{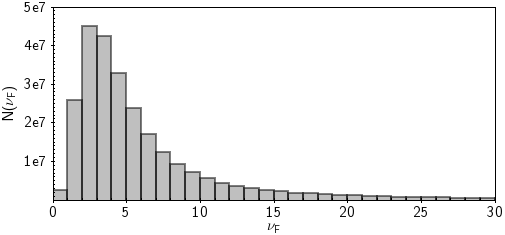}
	\caption{The distribution of $\nu_F$ in the z-band shear catalog.}
	\label{nu_f_dist}
\end{figure}

The spatial distribution of the galaxies in our final shear catalog (with the super-flat correction) is shown in fig.\ref{gal_density}. The typical galaxy number density is about 3 - 5 per square arcmin for z-band, and slightly less in the other two bands. The whole area of the distribution is more than ten thousand square degrees. In total, our catalog includes 99/111/116 million distinct galaxies in the band of g/r/z respectively. Since our shear measurement is performed on individual exposures independently, each galaxy has multiple shear estimators measured in each band due to the overlaps between the exposures. The distribution of the number of valid shear estimators for each galaxy are shown in fig.\ref{num_image_per_gal} for the three bands. The distribution of the selection function $\nu_F$ in the z-band catalog is shown in fig.\ref{nu_f_dist}.

\section{Conclusion and Discussions} 
\label{conclusion}

An interesting and potentially important question is about whether or not images of general seeing conditions are useful for accurate weak lensing measurement. In this work, we present our effort of answering this question by applying the Fourier\_Quad shear measurement pipeline on images taken in the DECaLS program, which is designed for locating targets of the DESI survey, not for weak lensing at all. The typical PSF FWHM in DECaLS is around 1.4-1.6 arcsec, much larger than those of the other existing weak lensing surveys, such as CFHTLenS\footnote{http://cfhtlens.org}, KiDS\footnote{https://kids.strw.leidenuniv.nl}, HSC\footnote{https://hsc.mtk.nao.ac.jp/ssp}, and DES\footnote{https://www.darkenergysurvey.org}.  

Our pipeline for this work is similar to the one introduced in \cite{Zhang2019} for processing the CFHTLenS data, albeit with significant modifications regarding PSF reconstruction, the details of which is given in \S\ref{PSF_recon}. This modification is necessary for improving the stability and accuracy of the identification of the PSF stars under poor seeing conditions. To achieve a fair quality in PSF reconstruction, we find it optimal to use fitting functions of the polynomial form up to the third order, although residuals of PSF ellipticities and sizes are still visible through stacking. The residual maps clearly reveal the tree ring features on the CCDs, consistent with those shown in \cite{Jarvis2021}.

Taking advantage of the large data volume of DECaLS, we try to construct the super-flat field for each CCD in every band. This step is introduced to our pipeline as a way of checking the image quality on small scales, rather than making accurate corrections. There turns out to be features such as dust attachment, funky columns, hot spots, and scratches showing up clearly, some of which are previously reported on the official website of the observatory. These problems are most serious for the g-band, and much milder for the z-band. 

These findings are consistent with the results of our field distortion test in \S\ref{fd}, which shows that the z-band shear catalog can most reliably recover the tiny field distortion signals ($\vert g\vert\le 0.0015$) for both $g_1$ and $g_2$. With the super-flat correction, the multiplicative and additive biases for the two shear components are all more or less consistent with zero at percent level: $m_1=0.05\pm 0.03$, $m_2=-0.03\pm 0.03$, $c_1=(0\pm 3)\times 10^{-5}$, $c_2=(-10\pm 3)\times 10^{-5}$. It is worth stressing that all these results are achieved without external calibrations. 

More interestingly, we divide the z-band galaxies into two groups according to whether or not the PSF FWHM is smaller than 1.4 arcsec. According to the results of the FD test shown in fig.\ref{FD_test_large_small_PSF}, the PSF size does not seem to affect the accuracy of the shear catalog, which is indeed very encouraging to know. 

Finally, we note some caveats in our current treatment of the DECaLS images. First, we have left out the discussion of the brighter-fatter effect \citep{Antilogus2014,Gruen2015}, which is caused by the charge interactions in the CCD, and can bias our shear measurement by rendering the PSF flux-dependent. The same physical process also introduces correlations between pixel noise, which is usually assumed to be Poissonian. In principle, removing these effects would require manipulations of pixel values for both stars and galaxies (see, e.g., \cite{Gruen2015}). The details though are out of the scope of this work. Second, we lack a way of treating the tree rings, which quite obviously stand out as the main contributor of the PSF residuals in fig.\ref{PSFsize_grz}. In a future work, we may consider using templates of the tree rings on individual CCDs to improve the modelling of the PSF (see, e.g., \cite{Plazas2014}). Lastly, the super-flat fields we construct from the images are so far mostly used for indicating small-scale problems in the CCDs. The corrections we make on the pixel level are still rough, and likely require more refinement.

Our shear catalogs in all three bands are now publicly available\footnote{https://gax.sjtu.edu.cn/data/DESI.html}. Our Fourier\_Quad image processing pipeline is also available by request.

\acknowledgments

This work is supported by the National Key Basic Research and Development Program of China (No.2018YFA0404504), and the NSFC grants (11621303, 11890691, 12073017), the science research grants from China Manned Space Project (No. CMS-CSST-2021-A01).
The computations in this paper were run on the $\pi$ 2.0 cluster supported by the Center for High Performance Computing at Shanghai Jiao Tong University.

The Legacy Imaging Surveys of the DESI footprint is supported by the Director, Office of Science, Office of High Energy Physics of the U.S. Department of Energy under Contract No. DE-AC02-05CH1123, by the National Energy Research Scientific Computing Center, a DOE Office of Science User Facility under the same contract; and by the U.S. National Science Foundation, Division of Astronomical Sciences under Contract No. AST-0950945 to NOAO.
The Photometric Redshifts for the Legacy Surveys (PRLS) catalog used in this paper was produced thanks to funding from the U.S. Department of Energy Office of Science, Office of High Energy Physics via grant DE-SC0007914.

\newpage
\bibliography{weaklensing}{}
\bibliographystyle{aasjournal}

\end{document}